\pdfoutput=1
\documentclass[%
 reprint,
 superscriptaddress,
 amsmath,amssymb,
 aps,
]{revtex4-2}

\usepackage[utf8]{inputenc}
\usepackage[T1]{fontenc}
\usepackage{graphicx}
\usepackage{dcolumn}
\usepackage{bm}
\usepackage{xurl}
\usepackage[hidelinks]{hyperref}

\begin{document}

\title{$\kappa$-Parametrized Boltzmann Distribution Sampling on Quantum Approximate Optimization Algorithm}

\author{João Marcos Cavalcanti de Albuquerque Neto}
\email{joao@niteqpuc.com}
\affiliation{NITeQ, Electrical Engineering Department, Pontifícia Universidade Católica do Rio de Janeiro, Rio de Janeiro 22451-900, Brazil}

\author{Gustavo Castro do Amaral}
\email{gustavo.castrodoamaral@tno.nl}
\affiliation{Quantum Technology Department, TNO---The Netherlands Organization for Applied Scientific Research, 2628CK Delft, The Netherlands}

\author{Guilherme Penello Temporão}
\email{temporao@opto.cetuc.puc-rio.br}
\affiliation{NITeQ, Electrical Engineering Department, Pontifícia Universidade Católica do Rio de Janeiro, Rio de Janeiro 22451-900, Brazil}

\begin{abstract}
Generative models use training data to produce new samples that resemble the original, with applications in classification and recommendation systems. Restricted Boltzmann Machines (RBMs) are at the core of generative models but suffer from the computational intractability of training, which relies on sampling an unknown Boltzmann Distribution (BD). The Quantum Approximate Optimization Algorithm (QAOA) can assist RBM training, potentially providing benefits over strictly classical strategies. One particular problem is determining the quantum circuit depth for an arbitrary problem size that yields the expected sampling results. In this study, we evaluate QAOA-assisted sampling of BDs using Qiskit's Aer Simulator by analyzing the inverse effective temperature of the distribution, $\kappa$, for tractable problem sizes. The results provide numerical evidence, based on Qiskit Aer simulations, that parameterized annealing-scheduled QAOA (PAQ-QAOA) can approximate RBM Boltzmann sampling for the tractable instances considered, while hardware performance and scaling remain open questions.
\end{abstract}


\maketitle

\section{Introduction}
Restricted Boltzmann Machines (RBMs) are probabilistic energy-based generative models used in machine learning. Since their introduction by Smolensky~\cite{smolensky1986information} and the development of efficient training methods by Hinton~\cite{hinton2002training}, RBMs have been applied to a wide range of tasks, including anomaly detection~\cite{fiore2013network}, recommendation systems~\cite{salakhutdinov2007restricted}, and financial-transaction fraud detection~\cite{neto2026fraud}. Using an energy-based formulation, RBMs can represent the joint distribution over their visible and hidden units using a Boltzmann distribution (BD).

Training an RBM classically is challenging: the goal is to match the distribution of the training data by adjusting weights of visible and hidden units, often via a log-likelihood function; the gradient of the latter can only be evaluated via sampling from a BD whose partition function is intractable for all but the smallest problem sizes. Classical approximation strategies, such as Gibbs Sampling (GS)~\cite{gelfand2000gibbs}, Contrastive Divergence (CD)~\cite{hinton2002training}, and Persistent Contrastive Divergence (PCD)~\cite{tieleman2008training}, provide workable heuristics that are not guaranteed to find the global optimum. These limitations have motivated the search for alternative sampling techniques that better approximate the model distribution while remaining computationally efficient. Quantum annealing (QA)~\cite{kadowaki1998quantum} has been explored for accelerating and improving BD sampling, since the non-adiabatic dynamics of quantum annealers can generate samples approximating a thermal distribution~\cite{amin2015searching}.

We investigate a deterministic framework for sampling BDs using the Quantum Approximate Optimization Algorithm (QAOA)~\cite{farhi2014quantum} with annealing parameterization~\cite{zhou2020quantum,sack2021quantum,sharma2022openqaoa}. This approach aims to mitigate the NP-hard optimization of QAOA parameters~\cite{bittel2021training} by replacing variational angles with a deterministic annealing schedule. Using numerical simulations spanning different circuit depths ($p$), time steps ($\Delta t$), and problem sizes, we identify three distinct sampling regimes and show that, among the regimes considered, the Non-Adiabatic Regime provides the closest approximation to the target BD. Sampling performance is evaluated using the Kullback--Leibler divergence~\cite{kullback1951information}. For small-qubit instances, the best sampling performance is observed near $\kappa \approx 1$, whereas deviations observed for larger systems suggest unresolved scaling effects.

\section{Restricted Boltzmann Machines}
\subsection{Definition}

RBMs are energy-based probabilistic generative models applicable to supervised and unsupervised learning tasks. An RBM consists of two layers: the visible and hidden layers. The first houses binary visible units $v_i$, which represent observed data, while the hidden layer contains binary hidden units $h_j$ responsible for capturing correlations among the visible variables. RBMs are energy-based models, and, therefore, their behavior is governed by an energy function:
\begin{equation}
E(\mathbf{v},\mathbf{h})
= -\mathbf{a}^{\top}\mathbf{v}
-\mathbf{b}^{\top}\mathbf{h}
-\mathbf{v}^{\top}\mathbf{W}\mathbf{h},
\label{eq:1}
\end{equation}
where the biases of the visible and hidden layers are denoted by $\mathbf{a}$ and $\mathbf{b}$, respectively, and $\mathbf{W}$ denotes the weight matrix connecting the two layers.

The energy function defines the joint probability distribution over the visible and hidden units in an RBM, which follows a BD,
\begin{equation}
p(\mathbf{v},\mathbf{h})=\frac{1}{Z}\,e^{-E(\mathbf{v},\mathbf{h})},
\qquad
Z=\sum_{\mathbf{v}}\sum_{\mathbf{h}} e^{-E(\mathbf{v},\mathbf{h})},
\label{eq:2}
\end{equation}
where $Z$ denotes the partition function associated with the BD.

By marginalizing over the hidden units, one obtains the probability distribution over the visible variables, which represent the data generated by the RBM:
\begin{equation}
    p(\mathbf{v}) = \frac{1}{Z}\exp\left( \mathbf{a}^{\top}\mathbf{v}+\sum_{j}\mathrm{softplus}\big(b_j+(\mathbf{W}^{\top}\mathbf{v})_j\big)\right).
    \label{eq:3}
\end{equation}
This relation demonstrates that by altering the biases and weights of the RBM, the probability distribution of its data will change accordingly.

\subsection{Training RBMs}
The training of an RBM aims to learn parameters $\theta=\{\mathbf{W},\mathbf{a},\mathbf{b}\}$ that maximize the data log-likelihood~\cite{hinton2002training},
\begin{equation}
\mathcal{L}(\theta)
= \sum_{n=1}^{N}\log p(\mathbf{v}^{(n)};\theta),
\end{equation}
thereby assigning higher probability to likely configurations, which correspond to lower-energy states. The model parameters can be optimized using stochastic gradient ascent (SGA)~\cite{hinton2012practical}, yielding the update rules
\begin{align}
\frac{\partial \mathcal{L}}{\partial \mathbf{W}}
&= \langle \mathbf{v}\mathbf{h}^{\top} \rangle_{\text{data}}
- \langle \mathbf{v}\mathbf{h}^{\top} \rangle_{\text{model}}, \\
\frac{\partial \mathcal{L}}{\partial \mathbf{b}}
&= \langle \mathbf{h} \rangle_{\text{data}} - \langle \mathbf{h} \rangle_{\text{model}}, \\
\frac{\partial \mathcal{L}}{\partial \mathbf{a}}
&= \langle \mathbf{v} \rangle_{\text{data}} - \langle \mathbf{v} \rangle_{\text{model}},
\end{align}
where the data-dependent expectations can be computed efficiently using the conditional activation probabilities of the hidden units,
\begin{equation}
p(h_j = 1 \mid \mathbf{v})
= \sigma\!\left(b_j + \sum_{i=1}^{N_v} W_{ij} v_i\right),
\end{equation}
whereas the model-dependent expectations require sampling from the model distribution, which is generally intractable due to the partition function $Z$~\cite{hinton2006fast}.

\subsection{Avoiding the Partition Function}
To date, there is no known computationally efficient method for exact sampling from Eq.~\eqref{eq:2}, motivating the development of approximate training methods. A common approach is to employ Gibbs Sampling (GS)~\cite{geman1984stochastic} to estimate the model-dependent expectations $\langle\cdot\rangle_{\text{model}}$ from $p(\mathbf{v},\mathbf{h})$. However, exact convergence to the target distribution is only guaranteed in the limit of an infinitely long Markov chain. To address this limitation, Hinton proposed Contrastive Divergence (CD)~\cite{hinton2002training}, which approximates the model-dependent expectations by running the Gibbs chain for a predetermined number of steps, $k$. A further refinement is Persistent Contrastive Divergence (PCD)~\cite{tieleman2008training}, which maintains a persistent Markov chain across parameter updates, rather than reinitializing the chain from the data at each iteration.

In summary, these methods provide approximations of the model expectations required for learning. Although such approximations do not yield exact samples from the model distribution, they can produce useful gradient estimates in practice~\cite{hinton2002training,hinton2012practical}. Moreover, improved estimators, such as PCD, have been shown to provide more reliable samples of the model distribution and more stable learning dynamics~\cite{tieleman2008training}.

\section{Quantum-Assisted RBM Training}
\subsection{Quantum Annealing}

QA was originally proposed as an approach to solving optimization problems~\cite{kadowaki1998quantum,farhi2000quantum}, based on the quantum adiabatic theorem~\cite{born1928}. This is done via the adiabatic time evolution of the ground state of a simple Hamiltonian ($H_1$) towards a complicated one ($H_2$):
\begin{equation}
    H(t) = \big(1-s(t)\big)H_1 + s(t)H_2,
    \label{eq:anneal}
\end{equation}
where $s(t)$ is a time-dependent annealing schedule responsible for the evolution of the instantaneous Hamiltonian $H(t)$, such that $s(0) = 0$ and $s(T) = 1$, where $T$ is the total annealing time. Under adiabatic evolution, the system remains in the instantaneous ground state of $H(t)$; consequently, the final state of the system will be the ground state of $H_2$~\cite{amin2015searching}.

When the evolution is too rapid to satisfy the adiabatic condition, the system enters a non-adiabatic regime, in which the final measurement distribution is no longer concentrated on the ground state of $H_2$. Instead, the system can approximately sample a thermal distribution of the form~\cite{amin2015searching,lotshaw2023approximate}
\begin{equation}
    P(\mathbf{x}) \propto e^{-\kappa H_2(\mathbf{x})},
    \label{eq:thermal}
\end{equation}
where $\kappa$ denotes the inverse effective temperature coefficient of the distribution. This non-adiabatic behavior is central to the sampling approach investigated in this study.

Notably, the adiabatic process proves useful in the context of RBM training because the energy function of an RBM can be written as a Quadratic Unconstrained Binary Optimization (QUBO) problem and, by extension, as an Ising Hamiltonian~\cite{lucas2014ising}:
\begin{equation}
    H_C = -\sum_{\langle i,j\rangle} J_{ij}\, Z_i Z_j - \sum_{i=1}^{n} g_i\, Z_i,
    \label{eq:Ising}
\end{equation}
where $J_{ij}$ encodes the coupling strength between qubits $i$ and $j$, and $g_i$ encodes the local field applied to qubit $i$. Since the RBM units take values in $\{0,1\}$ while the eigenvalues of $Z_i$ are $\pm1$, the mapping follows from the change of variables $x_i=(1-z_i)/2$, so that $J_{ij}$ is determined by the weights $\mathbf{W}$ and $g_i$ by a combination of the biases $\mathbf{a}$, $\mathbf{b}$ and the weights (Eq.~\eqref{eq:1}). In other words, by setting $H_1$ as a Hamiltonian with a trivial ground state and $H_2 = H_C$, quantum annealing can be used to approximately sample from the joint distribution associated with the RBM's energy function, assisting the gradient calculations in the training process~\cite{neto2026fraud}.

\subsection{Quantum Approximate Optimization Algorithm}

QAOA is a Variational Quantum Algorithm (VQA), originally proposed to solve combinatorial optimization problems~\cite{farhi2014quantum} and since demonstrated on quantum hardware~\cite{harrigan2021quantum}. VQAs make use of a parameterized quantum circuit with a classical optimization feedback. The goal of this hybrid quantum/classical feedback loop is to identify the quantum state that minimizes the expectation value of the Hamiltonian encoding the problem. From this perspective, QAOA and its variants are closely related to adiabatic quantum computing, except that they approximate continuous-time evolution through a discrete sequence of unitary operations. To explicitly make this connection, consider the time evolution under a Hamiltonian $H(t)$:
\begin{equation}
U(T)=\mathcal{T}\exp\left(-i\int_{0}^{T} H(t)\,dt\right).
\label{eq:12}
\end{equation}
By discretizing the time integral and applying the Trotter--Suzuki decomposition~\cite{suzuki1990fractal}, this time-evolution operator can be approximated as
\begin{equation}
U(T)\approx \prod_{k=1}^{p}
e^{-i\beta_k H_M}\,
e^{-i\gamma_k H_C},
\end{equation}
where $p$ represents the circuit depth, $H_M$ the mixing Hamiltonian with an easily prepared ground state, and $H_C$ the problem Hamiltonian. $\beta_k$ and $\gamma_k$ are parameters that, once optimized, allow sampling the ground state of $H_C$.

Notice that, similarly to QA, QAOA is used to sample the ground state of $H_C$ by ensuring that the process is adiabatic while controlling the circuit depth. Directly sampling a BD is challenging, although possible~\cite{lotshaw2023approximate}, and leads to deep circuits which, when implemented on NISQ computers, are limited by decoherence~\cite{preskill2018quantum}. Furthermore, as circuit depth increases, the classical optimization of $\beta_k$ and $\gamma_k$ becomes more challenging, to the point that the problem becomes intractable depending on size~\cite{bittel2021training}. One alternative is the introduction of an annealing schedule~\cite{zhou2020quantum,sack2021quantum,sharma2022openqaoa}, obtained by applying the Trotter--Suzuki decomposition to Eq.~\eqref{eq:12} with $H(t)$ given by Eq.~\eqref{eq:anneal}, $H_1=H_M$ and $H_2=H_C$:
\begin{equation}
U(T) \approx \prod_{k=1}^{p} e^{-i\Delta t\,(1-s_k)H_M}\, e^{-i\Delta t\, s_k H_C}.
\label{eq:annealQAOA}
\end{equation}
With this parameterization, the angles $\beta_k=(1-s_k)\Delta t$ and $\gamma_k=s_k\Delta t$ are fully determined by the annealing schedule $\{s_k\}$, eliminating the need for their intractable optimization. The remaining practical challenge is, therefore, to choose an appropriate circuit depth $p$ and time step $\Delta t$, which are discussed in the next section in the context of the parameterized annealing-scheduled QAOA (PAQ-QAOA) quantum circuit.

\subsection{PAQ-QAOA Quantum Circuit Definition}

\begin{figure}
    \centering
    \includegraphics[scale=0.19]{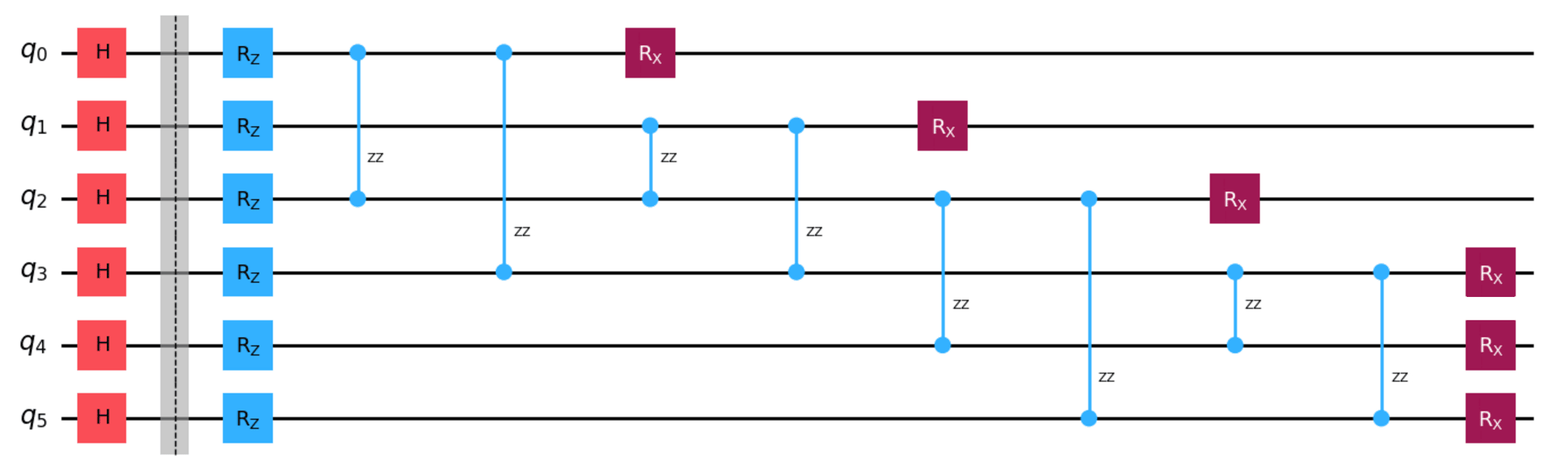}
    \caption{PAQ-QAOA quantum circuit for $p=1$, meaning a circuit depth of 1. In red, and before the gray bar, is the preparation of the ground state of $H_M$. In blue, the unitary generated by the Ising (problem) Hamiltonian $H_C$, and, in magenta, the unitary generated by the mixing Hamiltonian $H_M$.}
    \label{fig:circuit}
\end{figure}

Since $H_M$ is required to be a Hamiltonian with a trivial ground state, the transverse-field mixer $H_M=-\sum_{i=1}^{n} X_i$ is adopted. Its ground state is $|+\rangle^{\otimes n}$, which can be prepared by applying Hadamard gates to all qubits. Since the terms of $H_M$ commute,
\begin{equation}
    e^{-i\Delta t(1-s_k)H_M} = \bigotimes_{i=1}^{n} R_{X}^{(i)}\big( -2(1-s_k)\Delta t \big),
    \label{eq:15}
\end{equation}
with $R_X(\theta)=e^{-i\theta X/2}$. Similarly, since all terms of the Ising Hamiltonian of Eq.~\eqref{eq:Ising} commute,
\begin{equation}
    e^{-i \Delta t\, s_k H_C} = \prod_{\langle i,j\rangle} R_{ZZ}^{(ij)}\big(-2J_{ij}s_k\Delta t\big)\prod_{i=1}^{n}R_{Z}^{(i)}\big(-2g_i s_k\Delta t\big),
    \label{eq:16}
\end{equation}
with $R_{ZZ}(\theta)=e^{-i\theta Z\otimes Z/2}$ and $R_Z(\theta)=e^{-i\theta Z/2}$. By setting $s_k = k/p$, with $k=1,\dots,p$, the unitary corresponding to the PAQ-QAOA quantum circuit under the associated annealing schedule can be expressed as
\begin{equation}
\begin{split}
    U(p) = \prod_{k=1}^p \Bigg[ &\bigotimes_{i=1}^{n} R_X^{(i)}\Big( -2\big(1-\tfrac{k}{p}\big)\Delta t \Big) \\
    &\times \prod_{\langle i,j\rangle} R_{ZZ}^{(ij)}\Big(-2J_{ij}\tfrac{k}{p}\Delta t\Big)\prod_{i=1}^{n}R_{Z}^{(i)}\Big(-2g_i\tfrac{k}{p}\Delta t\Big) \Bigg].
\end{split}
\label{eq:17}
\end{equation}
The circuit representation of a single layer of this PAQ-QAOA is presented in Fig.~\ref{fig:circuit}.

\begin{figure*}
    \centering
    \includegraphics[scale=0.08]{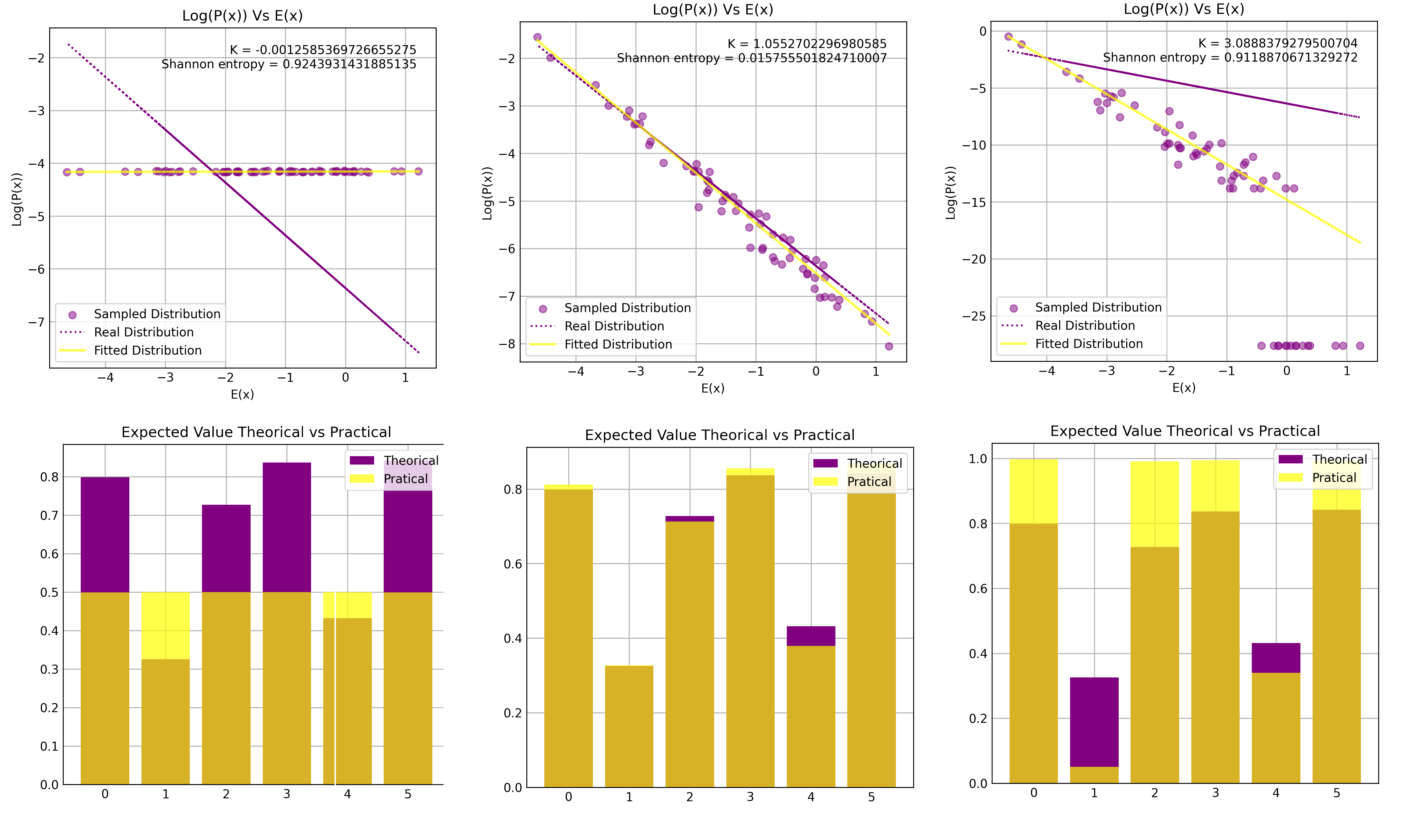}
    \caption{PAQ-QAOA sampling results in three different configurations of $p$ and $\Delta t$ with 6 qubits; parameters are manipulated so that the RR, NAR, and AR are reached. From left to right, these are: $p = 10$, $\Delta t = 0.121$; $p = 10$, $\Delta t = 0.28$; and $p = 10$, $\Delta t = 1.05$.}
    \label{fig:regimes}
\end{figure*}

\subsection{Time-Evolution Regimes}

Operation of PAQ-QAOA, although originally designed for RBM ground-state configuration search, reveals three distinct regimes associated with the choices of $p$ and $\Delta t$. For circuits whose depth is comparatively shallow with respect to the problem size, the sampled distribution resembles a uniform distribution; therefore, this regime is referred to as the \textit{Random Regime} (RR). Conversely, sufficiently deep circuits sample the ground state of $H_C$, as expected in standard QAOA. This corresponds to the solution of the original problem and is referred to as the \textit{Adiabatic Regime} (AR). For suitably tuned values of $p$ and $\Delta t$, which strongly depend on the problem instance, a regime can be reached in which the RBM's BD is sampled. Since the evolution in this configuration is faster than in the AR, this regime is defined as the \textit{Non-Adiabatic Regime} (NAR). The three different regimes, their corresponding sampled distributions, and the theoretical RBM's BD are shown in Fig.~\ref{fig:regimes}.

\subsection{The Inverse Effective Temperature $\kappa$}

To use PAQ-QAOA for sampling the RBM's BD, it is necessary to ensure, a priori, that the circuit parameters lead to the NAR. For small problem sizes, as illustrated in Fig.~\ref{fig:regimes}, calculation of the ground truth is tractable, allowing for parameter manipulation. However, this approach is not tractable for larger problem instances of practical interest. An alternative strategy is to relate the inverse effective temperature of the distribution, $\kappa$, to the relevant parameters of the PAQ-QAOA, namely the problem size $n$, $p$, and $\Delta t$. The motivation for this approach, as indicated by the results of Fig.~\ref{fig:regimes}, is that $\kappa$ consistently approaches unity when the PAQ-QAOA operates in the NAR. Note that, for hot distributions ($\kappa = 0$), sampling is random and the circuit operates in the RR; for cold distributions ($\kappa>1$), the circuit samples the ground state of $H_C$ as expected in the AR. The NAR occurs between these two regimes, with the RBM BD being most closely reproduced when $\kappa \approx 1$. This behavior can be evaluated by computing the KL divergence~\cite{kullback1951information} between the PAQ-QAOA sampled distributions in different regimes and the analytical ground truth for small problem sizes.

\begin{figure}
    \centering
    \includegraphics[scale=0.25]{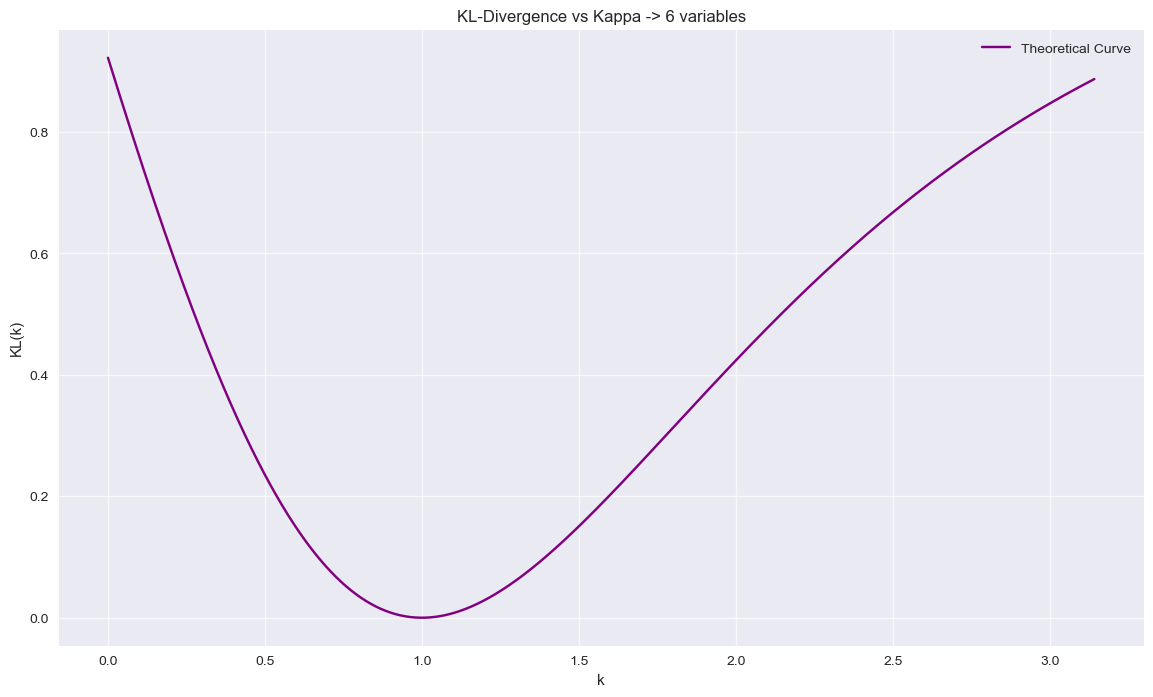}
    \caption{KL divergence as a function of the inverse temperature parameter $\kappa$, following Eq.~\eqref{eq:22}. As expected, for the small problem sizes where the calculation is tractable, $\kappa=1$ matches the lowest value of the KL divergence.}
    \label{fig:KLtheory}
\end{figure}

Assuming that $\kappa = 1$ provides an accurate indication that PAQ-QAOA operates in the NAR, the corresponding sampled probability distribution over the joint configurations $x_i=(\mathbf{v},\mathbf{h})$ can be modeled as
\begin{equation}
P_\kappa(x_i) = \frac{1}{Y(\kappa)}\, e^{-\kappa E(x_i)},
\label{eq:20}
\end{equation}
where $Y(\kappa)$ denotes the partition function associated with a given value of $\kappa$,
\begin{equation}
Y(\kappa) = \sum_{i} e^{-\kappa E(x_i)}.
\end{equation}
Inspection of Eq.~\eqref{eq:2} shows that $\kappa=1$ recovers exactly the BD of the RBM, with $Y(1)=Z$. Thus, within this model, $\kappa \approx 1$ indicates the desired BD sampling regime. Computing the KL divergence of the RBM's BD, Eq.~\eqref{eq:2}, from the model distribution of Eq.~\eqref{eq:20}, $D_{\mathrm{KL}}(P_\kappa\,\|\,p)=\sum_i P_\kappa(x_i)\log\left[P_\kappa(x_i)/p(x_i)\right]$, yields
\begin{equation}
\mathrm{KL}(\kappa) = \frac{1-\kappa}{Y(\kappa)} \sum_i E(x_i)\,e^{-\kappa E(x_i)} + \log\left(\frac{Z}{Y(\kappa)}\right),
\label{eq:22}
\end{equation}
which yields the curve presented in Fig.~\ref{fig:KLtheory}.

\section{Methodology and Results for the $\kappa$ Study}

\begin{figure*}
    \centering
    \includegraphics[scale=0.50]{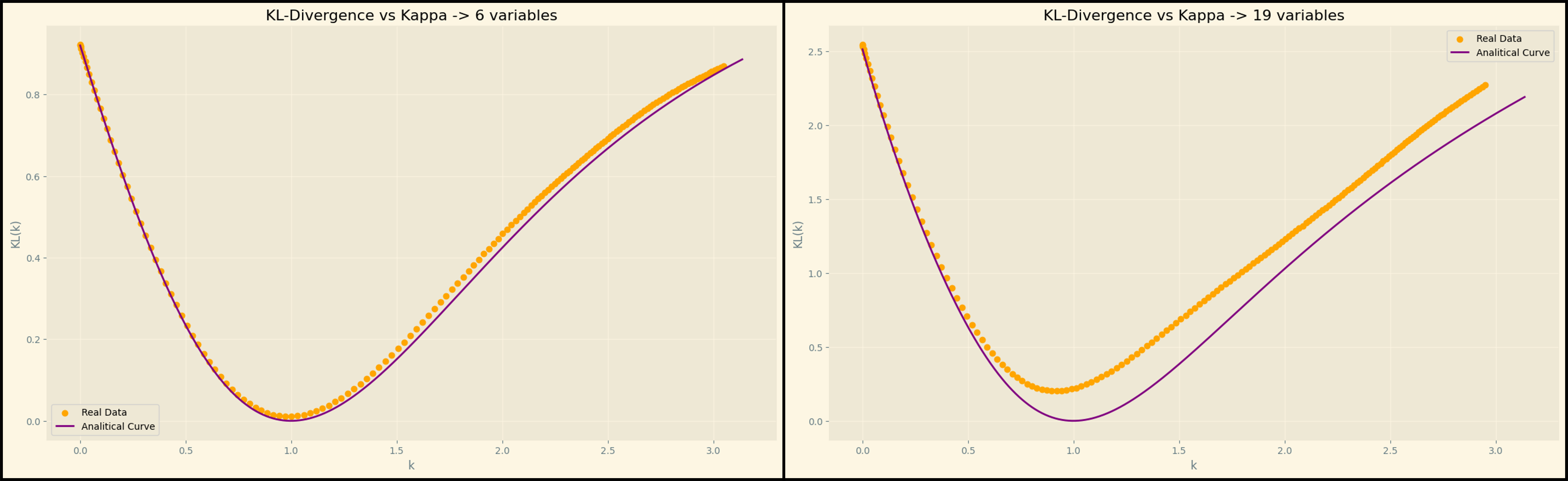}
    \caption{Qiskit Aer simulations of sampled and theoretical KL-divergence curves for $p=10$. For $n=6$ (left), the sampled curve closely follows the theoretical prediction, with the KL minimum at $\kappa \approx 1$. For $n=19$ (right), the sampled curve deviates from the theoretical prediction and its minimum shifts away from $\kappa = 1$.}
    \label{fig:results}
\end{figure*}

The main objective of this section is to provide a consistent methodology for evaluating whether the BD sampled by a PAQ-QAOA follows Eq.~\eqref{eq:20}. Such a methodology enables the identification of parameter configurations, for a given $n$ and selected values of $p$ and $\Delta t$, for which $\kappa = 1$ indicates sampling from the target BD. However, a model yielding $\kappa(n,p,\Delta t)=1$ is not yet available, since this relationship remains an open question. Therefore, the parameters are swept over a grid using Qiskit's Aer simulator. Results were obtained for a single RBM instance per problem size, and independent random seeds were used for initialization and sampling; while this prevents exact numerical reproducibility, key trends, such as the KL-divergence minimum near $\kappa \approx 1$, remained consistent across repeated runs. The steps to produce Fig.~\ref{fig:results} are:

\begin{enumerate}
    \item Theoretical KL-divergence calculation: RBM weights and biases are initialized from a uniform distribution over $[-1, 1]$. Using Eq.~\eqref{eq:22} for different values of $\kappa$, a curve analogous to Fig.~\ref{fig:KLtheory} is obtained.
    \item The RBM energy function is formulated as a QUBO problem, which is subsequently encoded into an Ising Hamiltonian.
    \item PAQ-QAOA circuit definition: the problem Hamiltonian, mixing Hamiltonian, $n$, $p$, $\Delta t$, and number of shots are computed and sent to Qiskit's Aer simulator.
    \item Sweep: the PAQ-QAOA circuit is executed for a range of $\Delta t$ values, and the corresponding sampled probability distributions are recorded. In this study, $\Delta t$ is sampled from a uniformly spaced grid over the interval $[0,\pi]$, consisting of 1,000 points, with $10^6$ shots performed for each value.
    \item $\kappa$ and KL-divergence calculation: for each value of $\Delta t$, $\kappa$ is estimated from the measured counts $N(x_i)$ via maximum likelihood estimation,
    \begin{equation}
    \kappa_{\text{est}}=\arg\max_{\kappa}\sum_i N(x_i)\log P_\kappa(x_i),
    \label{eq:mle}
    \end{equation}
    which is equivalent to matching the mean energy of the model distribution to the sampled mean energy; simultaneously, the KL divergence is calculated with the sampled distribution. This yields a pair $(\kappa_{\text{est}}, \mathrm{KL}_{\text{samp}})$, which is plotted against the theoretical curve for comparison.
\end{enumerate}

In order to assess problem tractability, two representative sizes were considered: $n=6$ (small) and $n=19$ (large but still tractable), with $p = 10$. For small problem sizes, the sampled and theoretical distributions closely match, and the KL divergence is minimized at $\kappa = 1$, supporting the expected correspondence with the RBM's BD. However, as the problem size increases, a divergence between the curves emerges, and the minimum of the KL divergence no longer occurs at $\kappa = 1$. This suggests that, while Eq.~\eqref{eq:22} captures a consistent relationship between $\kappa$ and the KL divergence, an unknown correction factor is still required.

\section{Conclusion}

This work investigated the use of PAQ-QAOA in the NAR to approximate sampling from an RBM Boltzmann distribution using Qiskit Aer simulations. The results indicate promising behavior for the simulated instances considered, while the practical determination of parameters leading to the NAR remains an open challenge. Future work includes analytical studies of the relationship between $\kappa$ and the KL divergence as a function of $n$, $p$, and $\Delta t$. Additionally, the statistical effects of NISQ hardware and the number of shots required for accurate sampling are still under investigation.

\section{Code Availability}

The source code used to generate the results reported in this study is publicly available in the
\href{https://github.com/NITeQ/--Parametrized-Boltzmann-Distribution-Sampling-on-Quantum-Approximate-Optimization-Algorithm}{project's GitHub repository}~\cite{pbds_qaoa}.
The repository includes routines for generating RBM parameters and provides additional reproducibility resources, including configurable random seeds, number of instances, and scripts for executing experiments with different circuit depths $p$, numbers of shots, and $\Delta t$ sweeps.

\begin{acknowledgments}
This study was financed in part by the Coordenação de Aperfeiçoamento de Pessoal de Nível Superior – Brasil (CAPES) – Finance Code 001, and by the Conselho Nacional de Desenvolvimento Científico e Tecnológico (CNPq), grant number 409596/2022-1. The authors also thank the members of the NITeQ group at PUC-Rio, especially A.~Wanick, C.~Paula, C.~Nascimento, G.~Santos, and M.~dos Magos.
\end{acknowledgments}

\bibliographystyle{ieeetr}
\bibliography{apssamp}

\end{document}